\documentclass{elsart}

\usepackage{psfig}
\usepackage{amsfonts}
\usepackage{amssymb}
\usepackage{amsbsy}
\usepackage{subeqn}

\newcommand{\Real}{\mathop{\mathrm{Re}}}
\newcommand{\Imag}{\mathop{\mathrm{Im}}}

\newcommand{\rmi}{{\rm i}}

\newcommand{\C}{\mathbb{C}}
\newcommand{\R}{\mathbb{R}}

\newcommand{\cK}{\mathcal{K}}
\newcommand{\cS}{\mathcal{S}}
\newcommand{\cC}{\mathcal{C}}

\newcommand{\bR}{\mathbf{R}}

\newcommand{\MeV}{\, {\rm MeV}}
\newcommand{\GeV}{\, {\rm GeV}}

\newcommand{\beq}{\begin{equation}}
\newcommand{\eeq}{\end{equation}}

\newcommand{\sds}{\strut\displaystyle}
\newcommand{\smallspace}{\vspace{2.5ex}}

\begin{document}
\begin{frontmatter}

\title{Time delay and time advance in resonance theory}

\author[EDM]{Enrico De Micheli}
\address[EDM]{IBF -- Consiglio Nazionale delle Ricerche, Via De Marini 6, 16149 Genova, Italy}
\ead{demicheli@ge.cnr.it}

\author[GAV]{Giovanni Alberto Viano}
\address[GAV]{Dipartimento di Fisica -- Universit\`a di Genova, Istituto
Nazionale di Fisica Nucleare, sez. di Genova, Via Dodecaneso 33,
16146 Genova, Italy}
\ead{viano@ge.infn.it}

\begin{abstract}
We propose a theory of the resonance--antiresonance scattering process
which differs considerably from the classical one (the Breit--Wigner theory),
which is commonly used in the phenomenological analysis. Here both
resonances and antiresonances are described in terms of poles
of the scattering amplitude: the resonances by poles in the first quadrant while
the antiresonances by poles in the fourth quadrant of the complex angular momentum plane.
The latter poles are produced by non--local potentials, which derive from the
Pauli exchange forces acting among the nucleons or the quarks composing the colliding particles.
\end{abstract}

\date{}

\end{frontmatter}

\section{Introduction}
\label{se:introduction}

The crucial assumption in the Breit--Wigner theory of resonances is the
causality condition. In its simplest form, it can be formulated as follows:
the outgoing wave cannot appear before the incoming wave has reached the
scatterer. If one assumes that the interaction is spherically symmetric,
linear, vanishing for $R>a$ ($a$ being the finite radius of the scatterer),
then it is sufficient to apply this causality condition at the surface of the scatterer
($R=a$) in order to guarantee causal propagation in the whole outside region.
Then, according to Eisenbud \cite{Eisenbud}, one can estimate the time delay (advance)
that the incident wave packet undergoes in the scattering process, by evaluating
the derivative of the phase--shift with respect to the energy. To be more
precise, we rapidly summarize the Wigner--Eisenbud analysis in the simplest case
of the $s$--wave phase--shift $\delta_0(E)$.

Using standard notations, we represent incoming and outgoing wave--packets
as \cite{Nussenzveig}
\begin{subequations}
\label{1}
\begin{eqnarray}
\psi_{\rm inc}(r,t) &=& \int_0^{+\infty} A(E)\,e^{-\rmi(kr+Et)}\,dE, \label{1a} \\
\!\!\!\psi_{\rm out}(r,t) &=& -\int_0^{+\infty} S(E)A(E)\,e^{\rmi(kr-Et)}\,dE=
-\int_0^{+\infty} A(E)\,e^{\rmi(kr+2\delta_0-Et)}\,dE, \label{1b}
\end{eqnarray}
\end{subequations}
where $S(E)=\exp[2\rmi\delta_0(E)]$. Let us suppose that $A(E)$
corresponds to a narrow energy spectrum, centered upon some energy $E_0$, so that
it takes appreciable values only for
\beq
|E-E_0| \lesssim \Delta E~~~~~(\Delta E \ll E_0).
\label{2}
\eeq
For large $|t|$ the integrands in Eqs. (\ref{1}) are rapidly oscillating
functions. Then the integrals in Eqs. (\ref{1}) can be evaluated by means of the stationary
phase method. Writing $A(E)=|A(E)|e^{\rmi\alpha(E)}$, the stationary phase point in
(\ref{1a}) can be recovered from the relation
\beq
\frac{d\alpha(E)}{dE}-r\frac{dk}{dE}-t = 0.
\label{3}
\eeq
Similarly, from (\ref{1b}) we have:
\beq
\frac{d\alpha(E)}{dE}+r\frac{dk}{dE}+2\frac{d\delta_0(E)}{dE}-t = 0.
\label{4}
\eeq
From (\ref{3}) and (\ref{4}) we can evaluate the time delay (or advance) between
the center of the incoming wave packet, which moves inward, and the center of the
outgoing wave packet, which moves outward; we have:
\beq
\Delta t = 2 \left(\frac{d\delta_0(E)}{dE}\right)_{E=E_0}.
\label{5}
\eeq
If we take for $\delta_0(E)$ the phase--shift due to the scattering by an
impenetrable sphere: i.e., $\delta_0(k)=-ka$ ($a=$ radius of the sphere),
and adopt appropriate units which allow us to write $E=k^2/2$, then
from (\ref{5}) we obtain:
\beq
\Delta t = -\frac{2a}{k}.
\label{6}
\eeq
It corresponds to a path difference $2a$ (from the surface to the center and back
of a sphere of radius $a$), between a wave reflected at the surface of the sphere and a
wave passing through its center. Correspondingly, an outgoing signal can appear at a time
earlier than it would have been possible in absence of the scatterer.

Returning to formula (\ref{5}), Wigner \cite{Wigner} has given a physical interpretation of this
result in terms of causality. The incident wave packet can be captured by the scatterer and
retained for an arbitrarily long time, so that there is no upper bound for the time delay.
Conversely, causality does not allow an arbitrarily large negative delay (time advance).
Classically, the maximum time advance allowed is that given by formula (\ref{6}).
Additional corrective terms of the type $1/k$, which are of the
order of a wavelength, would arise from the wave nature of matter \cite{Nussenzveig}.

If we transfer the results of this analysis to the representation of amplitudes and
cross--sections then we must take into account two types of contributions:
a pole singularity (resonance contribution) and the so--called hard sphere scattering contribution.
It turns out that the experimental phase--shift $\delta_0(E)$ can be reproduced by
patching up two pieces: $\delta_0^{(\rm res)}(E)$ and $\delta_0^{(\rm pot)}(E)$
(see formulae (\ref{new1}) in Section \ref{se:2}). More specifically:

(i) {\it Resonance contribution}: $S^{(\rm res)}(E)|_{E \in U} = e^{2\rmi\delta_0^{(\rm res)}(E)}$,
where $U$ is a neighborhood of the resonance energy. This term represents the approximation of the
scattering function $S(E)$, which is appropriate in a domain close to the resonance energy.
The function $S(E)$, regarded as a function of $E$,
presents a two--sheeted Riemann surface, corresponding to $k=(2E)^{1/2}$. In the
first sheet $S(E)$ is regular, except for possible poles on the negative
real axis, which correspond to the bound states energies. On the second (unphysical)
sheet, $S(E)$ presents complex poles in the neighborhood of the resonances, which
always arise in complex conjugate pairs (corresponding to $k$ and $-k^*$). One can regard
the poles in the fourth quadrant of the complex $k$--plane as associated with {\it decaying states},
and the poles in the third quadrant as corresponding to {\it capture states} \cite{Nussenzveig}.
The square of the imaginary
part of the location of these poles is related to the width of the resonance, and it is
inversely proportional to its lifetime, then to the time delay.

(ii) {\it Hard--sphere scattering contribution}: The term
$S^{(\rm pot)}(k)= e^{2\rmi\delta_0^{(\rm pot)}(k)}= e^{-2\rmi ka}$
is related to the so--called potential scattering, which
is responsible, according to formula (\ref{6}), for the time advance
(see also formulae (\ref{2.5})--(\ref{new1}) in Section \ref{se:2}).

The above discussion is mainly qualitative, and other definitions of the time delay
have been proposed; the interested reader is referred to Ref. \cite{Nussenzveig} for an exhaustive
review of these theories. Particularly significant is that proposed by Smith \cite{Smith}, and
extended by Goldberger and Watson \cite{Goldberger1} (see also Ref. \cite{Goldberger2}). In this theory, instead
of formula (\ref{5}), one obtains the following one:
\beq
\langle \Delta t \rangle = \left\langle 2\frac{d\delta_0(E)}{dE}\right\rangle_{\rm in},
\label{7}
\eeq
where the right hand side denotes the expectation value of $2\frac{d\delta_0(E)}{dE}$ in the
initial state (incoming wave packet); this expectation value gives the average time delay
due to the interaction. In particular, if the energy spectrum of the initial state is
centered around $E=E_0$, and is sufficiently narrow, then one recovers formula (\ref{5}) again. We
can thus say that all these proposals can be viewed as variations on a single theme, the causality
principle remaining the milestone. Therefore, hereafter we shall refer, for simplicity, to the
Breit--Wigner theory only.

We now want to raise two issues:
\begin{itemize}
\item[(i)] Can the time advance be evaluated in terms of hard--core scattering
in those collisions between composite particles, when Pauli exchange forces enter the game?
\item[(ii)] Time delay and time advance are not described in a symmetrical way: the former is evaluated
from the pole singularities of the scattering amplitude; the latter by the scattering from an impenetrable sphere.
Then a clear--cut separation between these two terms is missing, and their interference cannot
be easily controlled.
\end{itemize}
Hereafter we present a theory whose peculiar character consists in evaluating both time delay
and time advance by means of scattering amplitude singularities (poles), which lie, respectively,
in the first and fourth quadrant of the complex angular momentum (CAM) plane. In this type of
approach to scattering phenomena, the resonances are grouped in families and lie
on trajectories of the angular momentum, regarded as a function of the energy.
Then, one has to face the problem of connecting the resonances belonging to an ordered family, the order being
given by the value of the angular momentum: the point is that
between two resonances of the same family there is the so--called
{\it echo} of the resonance \cite{McVoy}, or {\it antiresonance}, which is responsible for the time advance.
Then a feasible description of a family of resonances should account also for the related antiresonances.
Although the term antiresonance can be misleading since it could evoke the concept of {\it antiparticle},
and this is not the case indeed, we shall speak frequently of antiresonance in order to
emphasize its deep connection with the resonance and, in what follows, the terms echo of the resonance and
antiresonance will be used interchangeably.

To be more specific, let us consider the collision between two clusters (like $\alpha$--$\alpha$ scattering)
or two composite particles (like $\pi^+$--p scattering); when beyond a resonance we observe
an antiresonance, the latter can be due to the Pauli exchange forces which arise whenever
the interacting particles penetrate each other and the fermionic character of the
components (i.e., nucleons or quarks) emerges. At this point, however, we strongly remark that
there does not exist a one--to--one correspondence between time advance, compositeness of the
interacting particles and Pauli exchange forces, as we shall explain with more details and examples
in a remark in Section \ref{se:3}. Coming back to the Pauli exchange forces, we note that it is
precisely the Pauli antisymmetrization which leads us to introduce
non--local potentials \cite{DeMicheli2}, which depend on
the angular momentum. Then performing for this enlarged class of potentials the analytic continuation
of the partial scattering amplitudes from integer values of the angular momentum $\ell$
to complex--valued angular momenta $\lambda \in \C$, a peculiar feature, which is
absent in the case of local Yukawian potentials, comes out: the scattering amplitude can have
pole singularities both in the first and in the fourth quadrant of the CAM plane.
The poles located in the first quadrant (i.e., $\Real\lambda \geqslant 0,\,\Imag\lambda > 0$)
correspond to resonances (i.e., unstable states) and the time delay is related to $\Imag\lambda >0$;
instead, the poles lying in the fourth quadrant ($\Real\lambda\geqslant 0$, $\Imag\lambda <0$)
cannot be related in any way to unstable state: they correspond to the echoes of the
resonances (i.e., the antiresonances). In this type of approach both time delay
and time advance are described in terms of scattering amplitude singularities (poles), which, however,
{\it act at different values of energy}: when the resonance pole is dominant, the pole
corresponding to the echo can be neglected, and vice versa. Therefore, in the
present theory we can answer the previous questions in the following sense:
\begin{itemize}
\item[(i')] When clusters of particles or composite particles penetrate each other in the collisions
producing antiresonances, and these latter are due to Pauli exchange forces, then the echoes of the
resonances can be described by the use of pole singularities associated with non--local,
angular momentum dependent  potentials, which are generated by the Pauli exchange forces.
\item[(ii')] Time delay and time advance are treated symmetrically since both are described by the use of
scattering amplitude poles which act at different values of the energy: in the range
of energy where a resonance pole is dominant the effect of the corresponding antiresonance pole
is negligible, and vice versa.
\end{itemize}

The literature on the concepts of resonance, time delay, and on the methods for
extracting the resonance parameters from the experimental data is enormous,
and it is growing up in recent years (see, for instance, Refs.
\cite{Dalitz2,Cutkosky,Vrana,Bransden,Pelaez,Svenne,Kelkar3} and the references therein).
Among these many works, we limit ourselves to quote a quite recent approach which seems
promising and consists in extracting the resonance parameters from the energy distribution
of the time delay \cite{Kelkar,Kelkar2}.

The paper is organized as follows. In Section \ref{se:2} we present some phenomenological
examples which show the difficulties connected with the evaluation of the time delay when the
standard approach is used. In Section \ref{se:3} the non--local potentials are introduced, and
we show how the echoes of the resonances can be explained by introducing the Pauli exchange forces
which generate non--locality. In Section \ref{se:4} the representation of
resonances and antiresonances in the CAM plane is treated, and, finally, in
Section \ref{se:5} the theory is tested on the $\alpha$--$\alpha$ and $\pi^+$--p elastic scattering.

\section{Time delay and time advance in the resonance--antiresonance theory}
\label{se:2}
\begin{figure}[t]
\begin{center}
\leavevmode
\psfig{file=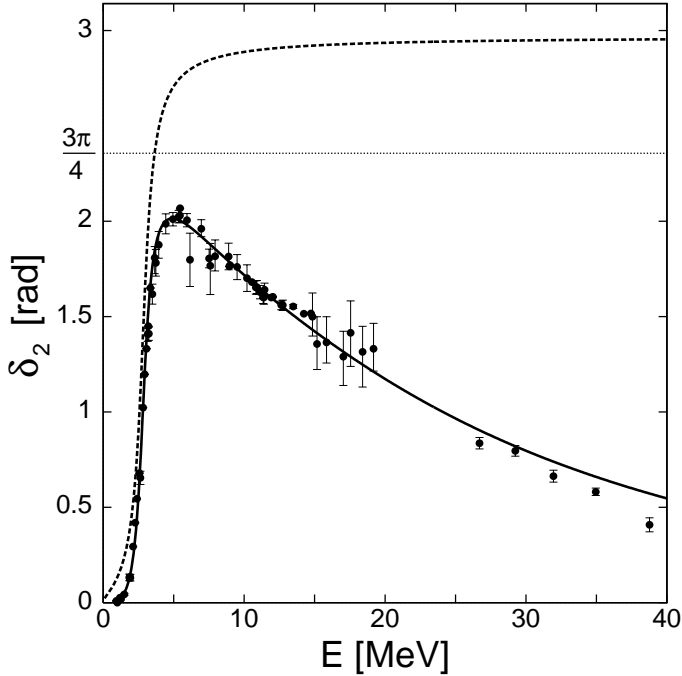,width=9cm}
\caption{\label{fig_1} \baselineskip=13pt $\alpha$--$\alpha$ elastic scattering.
Experimental phase--shifts (dots) for the partial wave $\ell=2$ and
corresponding fits vs. the center of mass energy $E$.
The experimental data are taken from Refs. \protect\cite{Afzal,Buck,Chien,Darriulat,Tombrello}.
The solid line indicates the phase--shift computed by using the symmetrized form of
formula (\protect\ref{4.13}), which takes into account both resonance and antiresonance terms.
The dashed line shows the phase--shift computed by using only the resonance term (see formula
(\protect\ref{4.10})).
The numerical values of the fitting parameters are (see Section \ref{se:5}):
$I=0.76$ (MeV)$^{-1}$, $\alpha_0=1.6$, $b_1=1.06\times 10^{-1}$ (MeV)$^{-1/2}$,
$a_1=1.03$ (MeV)$^{-1/4}$, $g_0=0.72$, $g_1=-7.5\times 10^{-3}$ (MeV)$^{-1}$,
$g_2=2.0\times 10^{-5}$ (MeV)$^{-2}$, $E_*=4.1$ MeV.}
\end{center}
\end{figure}

\begin{figure}[ht]
\begin{center}
\leavevmode
\psfig{file=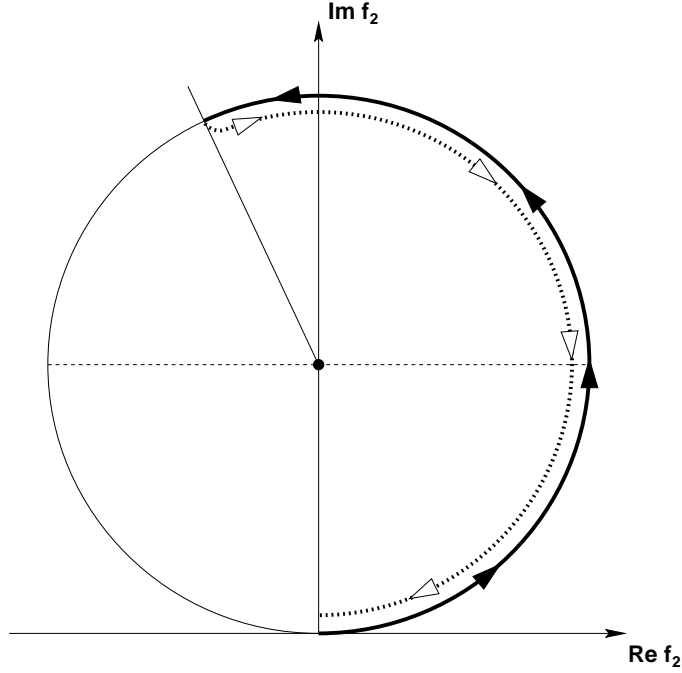,width=9cm}
\caption{\label{fig_2} \baselineskip=13pt $\alpha$--$\alpha$ elastic scattering.
Argand plot for the partial wave $f_{(\ell=2)}$ as a function of the energy.}
\end{center}
\end{figure}

Let $f_\ell = (e^{2\rmi\delta_\ell}-1)/(2\rmi)$ denote the partial wave amplitude of angular
momentum $\ell$, $\delta_\ell$ being the phase--shift, and assume that the condition of elastic
unitarity holds true (see the remark below). If the partial wave $f_\ell$ is represented as a vector
in the complex plane, causality requires that, as the energy increases, the vector will trace a circle in
counterclockwise sense. The vector describes a semi--circle of radius $\frac{1}{2}$
and center $\frac{\rmi}{2}$ as $\delta_\ell$ varies from $0$ to $\frac{\pi}{2}$; when
$\delta_\ell=\frac{\pi}{2}$, $f_\ell=\rmi$. If, for increasing energy,
$\delta_\ell$ tends to $\pi$, then this circle will close. In this case the width
$\Gamma$ of the resonance can be determined as the difference in energy at the opposite ends of the diameter
parallel to the real axis. It is clear that this method can work only if $\delta_\ell$
increases, for increasing energy, at least up to $\delta_\ell=\frac{3}{4}\pi$; in fact,
to this value of $\delta_\ell$ there corresponds the partial wave $f_\ell=-\frac{1}{2}+\frac{\rmi}{2}$,
and the distance in energy between the points $(-\frac{1}{2}+\frac{\rmi}{2})$ and $(\frac{1}{2}+\frac{\rmi}{2})$
can be determined. If $\delta_\ell$ does not reach $\frac{3}{4}\pi$, the method cannot be applied.
This is, for instance, the case of the phase--shift $\delta_{(\ell=2)}$ generated in the
$\alpha$--$\alpha$ elastic scattering (see Fig. \ref{fig_1}). Let us focus
on this example and plot in Fig. \ref{fig_2} the vector representing the partial
wave $f_{(\ell=2)}$ as a function of $E$. This vector,
which describes a circular arc with radius $\frac{1}{2}$ and center $\frac{\rmi}{2}$, moves
in counterclockwise sense up to the energy $E \simeq 4.9 \MeV$;
correspondingly, $\delta_{(\ell=2)}$ reaches its maximum value, $\delta_{(\ell=2)} \simeq 2.01$.
After, for increasing energy, $\delta_{(\ell=2)}$ decreases, passing downward through $\frac{\pi}{2}$;
consequently the plot of $f_{(\ell=2)}$ describes the same circular arc, but now covered in clockwise sense.
At $E \simeq 12.4 \MeV$ we have $\delta_{(\ell=2)} = \frac{\pi}{2}$, and $f_{\ell=2}=\rmi$: we have the antiresonance.
For $E > 12.4 \MeV$, $\delta_{(\ell=2)}$ keeps decreasing, approaching zero; accordingly, the vector representing
$f_{(\ell=2)}$ moves in clockwise sense toward the origin. From this phenomenological example it can manifestly be
seen that the above method cannot be used to determine the width $\Gamma$, in this specific case.

\smallspace

\noindent
{\bf Remark.} We want to stress the fact that our analysis is strongly focused on the pure elastic scattering region.
This choice is motivated by the consideration that in order to understand more clearly the
role of the antiresonances, it is convenient to isolate the echoes of the resonances
from other effects related to inelasticity.

\smallspace

Let us now write the partial wave amplitude as $f_\ell=e^{\rmi\delta_\ell}\sin\delta_\ell=\frac{1}{\cot\delta_\ell-\rmi}$,
and expand $\cot\delta_\ell(E)$ in Taylor's series in the neighborhood of the resonance, whose
center of mass energy is denoted by $E_r$. For energies close to the resonance $\delta_\ell \simeq \frac{\pi}{2}$,
and $\cot\delta_\ell(E) \simeq 0$ ($E$ denotes the total energy of the two particle state in the
center of mass system). Then we have:
\beq
\!\!\!
\cot\delta_\ell(E)\simeq\cot\delta_\ell(E_r)+(E-E_r)\left[\frac{d}{dE}
\cot\delta_\ell(E)\right]_{E=E_r} \simeq -\frac{2}{\Gamma}(E-E_r),
\label{2.1}
\eeq
where we have defined
$\frac{2}{\Gamma}=\left(\frac{d\delta_\ell(E)}{dE}\right)_{E=E_r}$. Additional terms in the series
can be neglected as far as $|E-E_r| \simeq E_r$, i.e., when the width of the resonance is small compared to
its energy. Then we have:
\beq
f_\ell(E)=\frac{1}{\cot\delta_\ell(E)-\rmi}=
\frac{\frac{\Gamma}{2}}{(E_r-E)-\rmi\Gamma/2}.
\label{2.2}
\eeq
Let us now recall the expression of the total cross--section for elastic scattering:
\beq
\sigma_{\rm el} = \frac{4\pi}{k^2}\sum_{\ell}(2\ell+1)\sin^2\delta_\ell=
\frac{4\pi}{k^2}\sum_{\ell}(2\ell+1)\left|\frac{e^{2\rmi\delta_\ell}-1}{2\rmi}\right|^2.
\label{2.3}
\eeq
Assuming that near the resonance all the phase--shifts $\delta_\ell$ are zero but one,
from (\ref{2.2}) and (\ref{2.3}) we obtain:
\beq
\sigma_{\rm el} = \frac{4\pi}{k^2} (2\ell+1) \frac{\Gamma^2/4}{(E-E_r)^2+\Gamma^2/4},
\label{2.4}
\eeq
which is the well--known Breit--Wigner formula for a resonant cross--section.
The resonance curve $\sigma_{\rm el}(E)$ is symmetric around
$E=E_r$, and the width $\Gamma$ is defined such that the cross--section at $|E-E_r|=\frac{\Gamma}{2}$ is half
of its maximum value.

\begin{figure}[t]
\begin{center}
\leavevmode
\psfig{file=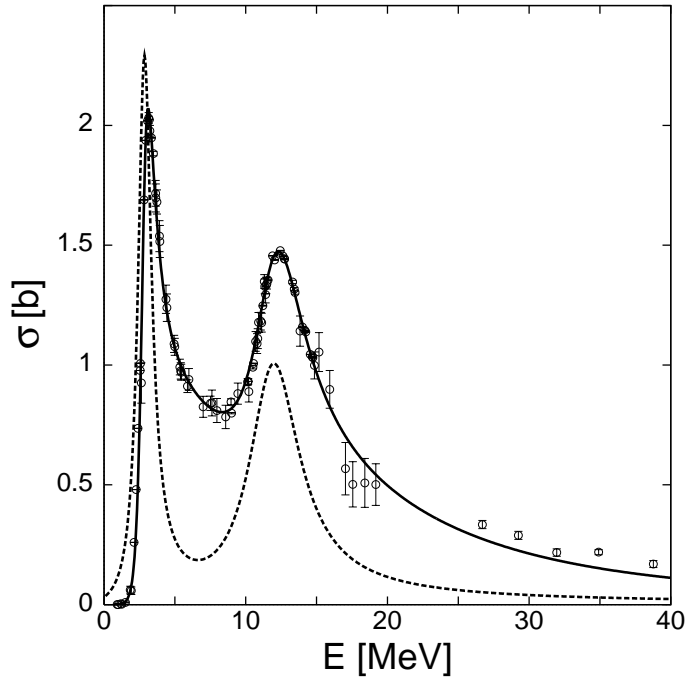,width=9cm}
\caption{\label{fig_3} \baselineskip=13pt $\alpha$--$\alpha$ elastic scattering.
Comparison between the total cross--section $\sigma(E)$ (see Eq. (\protect\ref{4.5}))
computed by accounting for both
the resonance and antiresonance terms (solid line) and that computed by using
only the resonance term (dashed line).}
\end{center}
\end{figure}

\begin{figure}[t]
\begin{center}
\leavevmode
\psfig{file=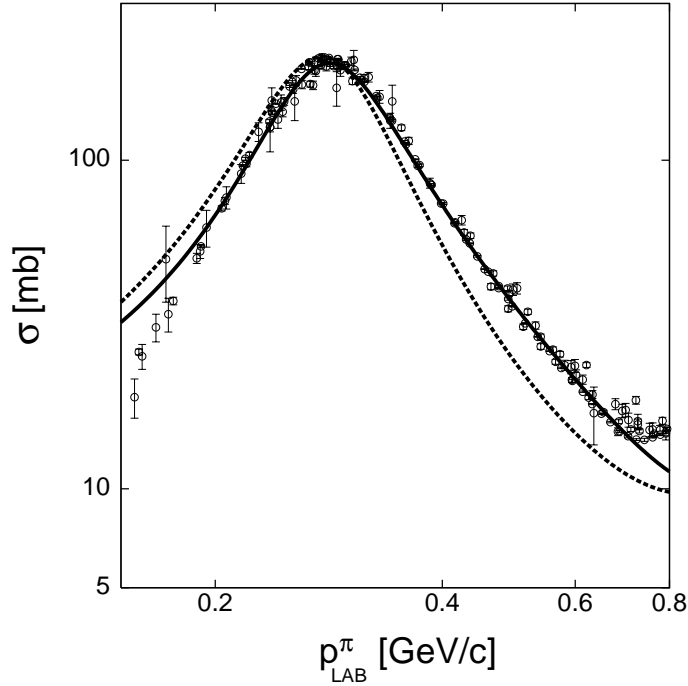,width=9cm}
\caption{\label{fig_4} \baselineskip=13pt $\pi^+$--p elastic scattering:
total cross--section as a function of the laboratory beam momentum. The experimental data (dots) are taken from
Ref. \protect\cite{Hagiwara}. The solid line indicates the total cross--section
computed by taking into account the
contributions of both the resonance and antiresonance poles generating
$\delta_\ell^{(+)}$ (formula (\protect\ref{4.21})).
The dashed line shows the total cross--section computed by accounting only
for the resonance poles generating $\delta_\ell^{(+)}$.
The fitting parameters are (see (\protect\ref{4.23})):
$a_0=6.95\times 10^{-1}$,
$a_1=9.0\times 10^{-7}$ (MeV)$^{-2}$,
$b_1=1.0\times 10^{-4}$ (MeV)$^{-1}$,
$b_2=1.4\times 10^{-7}$ (MeV)$^{-2}$, $c_0=-0.5$,
$c_1=5.0\times 10^{-7}$ (MeV)$^{-2}$,
$g_0=2.0\times 10^{-6}$ (MeV)$^{-2}$, $g_1=5.0\times 10^{-12}$ (MeV)$^{-4}$.}
\end{center}
\end{figure}

But, in Fig. \ref{fig_3}, where the plot of the total $\alpha$--$\alpha$ elastic scattering
cross--section against the center of mass system energy is shown, we clearly see that the two resonance
peaks, which correspond to $\ell=2$ and $\ell=4$, are clearly asymmetric around their resonance energies
$E^{(\ell=2)}_r=3.23 \MeV$ and $E^{(\ell=4)}_r=12.6 \MeV$. These asymmetries are produced precisely by
the antiresonances, which correspond to the downward passage of $\delta_{(\ell=2)}$ and $\delta_{(\ell=4)}$
through $\frac{\pi}{2}$. Indeed, the antiresonances do not produce sharp peaks but an
asymmetric fall in the elastic total cross--section. A similar effect appears also
in Fig. \ref{fig_4}, which refers to the $\Delta(\frac{3}{2},\frac{3}{2})$ resonance
in the $\pi^+$--p elastic scattering.

In order to remedy this defect in the Breit--Wigner theory, to the resonant partial wave $f_\ell$
in (\ref{2.2}) it is added the contribution produced by the so--called potential scattering,
whose amplitude is the same as the one for the scattering from an impenetrable sphere: i.e., hard--core scattering.
Thus, instead of formula (\ref{2.4}), the data are fitted by using the following formula \cite{Blatt}:
\beq
\sigma_{\rm el} = \frac{4\pi}{k^2}(2\ell+1)\left|\frac{\Gamma/2}{(E_r-E)-\rmi\Gamma/2}
+f_\ell^{\rm pot}\right|^2,
\label{2.5}
\eeq
where $f_\ell^{\rm pot}$ denotes the partial wave amplitude for the scattering from an impenetrable
sphere, and whose corresponding phase--shifts are given by \cite{Mott}:
\beq
\delta_\ell^{\rm pot}=\tan^{-1}\frac{J_{\ell+1/2}(ka)}{N_{\ell+1/2}(ka)},
\label{2.6}
\eeq
where $a$ is the radius of the sphere, while $J_{\ell+1/2}(\cdot)$ and $N_{\ell+1/2}(\cdot)$ denote the Bessel and Neumann
functions of index $(\ell+\frac{1}{2})$, respectively.
The asymptotic formulae representing $J_{\ell+1/2}(ka)$ and $N_{\ell+1/2}(ka)$ for large values of
$k$ read \cite{Tikhonov}:
\begin{subequations}
\label{2.7}
\begin{eqnarray}
J_{\ell+1/2}(ka) &=& \left(\frac{2}{\pi ka}\right)^{1/2}\cos\left[ka-\left(\ell+\frac{1}{2}\right)
\frac{\pi}{2}-\frac{\pi}{4}\right] + O\left((ka)^{-3/2}\right), \label{2.7a} \\
N_{\ell+1/2}(ka) &=& \left(\frac{2}{\pi ka}\right)^{1/2}\sin\left[ka-\left(\ell+\frac{1}{2}\right)
\frac{\pi}{2}-\frac{\pi}{4}\right] + O\left((ka)^{-3/2}\right). \label{2.7b}
\end{eqnarray}
\end{subequations}
Eqs. (\ref{2.6}) and (\ref{2.7}) yield the following asymptotic formula representing
$\delta_\ell^{\rm pot}(k)$ for large values of $k$:
\beq
\delta_\ell^{\rm pot}(k) = -ka + \frac{\pi}{2}\ell + O\left((ka)^{-3/2}\right).
\label{2.8}
\eeq
From (\ref{2.8}) it follows that for those values of $k$ such that $ka=(\ell+1)\frac{\pi}{2}$,
then $\delta_\ell^{\rm pot}(k) \simeq -\frac{\pi}{2}$ (${\rm mod}\,\pi$), and we have an antiresonance.

In the classical theory presented so far the phase--shift $\delta_\ell(E)$ is
then fitted by adding two contributions:
\begin{subequations}
\label{new1}
\begin{eqnarray}
\delta_\ell^{(\rm res)} &=& \tan^{-1} \left(\frac{\Gamma/2}{E_r-E}\right), \label{new1a} \\
\delta_\ell^{(\rm pot)} &=& -ka + \frac{\pi}{2}\ell, \label{new1b}
\end{eqnarray}
\end{subequations}
the approximation (\ref{new1a}) being feasible for sufficiently narrow resonance, and formula (\ref{new1b})
holding for sufficiently high values of $ka$. However, these last two constraints are not sufficient to
justify formula (\ref{2.5}). In fact, the step from formula (\ref{2.3}) to (\ref{2.5}) is admissible only if
in the energy range in which the resonance term is dominant, i.e., $E \simeq E_r$, the contribution
due to the potential scattering is negligible, and also vice versa, when the potential scattering effect is
dominant the resonance term has to be negligible.

Now, let us come back to formulae (\ref{2.1})--(\ref{2.4}). They represent what could be called the
{\it orthodox} Breit--Wigner formalism. If rigidly applied, these formulae would describe appropriately
the phase--shifts only near the resonance energy. The threshold is not correctly described and neither
is the high energy part. A correct description of the threshold can be achieved by using a
phase--space corrected width rather than a constant one. This amounts to introduce an $\ell$ dependent
threshold factor to guarantee that the cross--section and the phase--shifts go smoothly to zero
at threshold, instead of approaching finite constants. Now, if the threshold is correctly described,
nevertheless it can happen that on the high energy side the description is worse: the width keeps
growing with energy. Then the high energy behavior can be phenomenologically corrected by introducing appropriate
form factors, which account also for the fact that the interacting particles are not pointlike.
These corrections are widely applied in the phenomenological analysis \cite{Cutkosky,Arndt0,Arndt,Manley}.

We shall follow another procedure: we still introduce the antiresonances; however, the echoes will not
be described by $\delta_\ell^{(\rm pot)}$ (as in formula (\ref{new1b})) but by the contribution to
$\delta_\ell$ due to a pole singularity. More precisely, and this is the relevant point in the
theory we present, both contributions to $\delta_\ell$, due to the resonance and the antiresonance,
are described in terms of pole singularities,
lying, respectively, in the first and fourth
quadrant of the CAM plane, {\it but acting at different values of the energy}.
In the neighborhood of a resonance the resonance pole is dominant and the effect of the
antiresonance singularity can be neglected; conversely, at energies close to the antiresonance the
corresponding pole produces a large contribution while the resonance pole is negligible.

\section{Antiresonances and non--local potentials}
\label{se:3}
Let us first consider the scattering from a local central potential with finite first and second
moments. If we also assume that the potential does not have an $s$--wave bound state
at zero energy, then the standard Levinson's equality for the phase--shift $\delta_\ell(k)$ reads
as follows \cite{DeAlfaro}:
\beq
\delta_\ell(0)-\delta_\ell(\infty) = N_\ell \pi,
\label{3.1}
\eeq
where $N_\ell$ is the number of $\ell$--wave bound states. In the case of non--local potentials,
equality (\ref{3.1}) must be modified in order to take into account also the spurious bound
states, that is, bound states with positive energy \cite{Bertero}. For the sake of simplicity, and without loss of
generality, we neglect this peculiar feature of the non--local potentials, and assume that the
potential being considered does not present bound states embedded in the continuum.
Therefore we maintain equality (\ref{3.1}), which, rewritten in terms of the lifetime
$\tau_\ell(E)=2\frac{d\delta_\ell(E)}{dE}$, reads\footnote{In the proof of Levinson's theorem
one does not go from formula (\ref{3.1}) to equality (\ref{3.2}) but vice versa from (\ref{3.2})
to (\ref{3.1}). Indeed, formula (\ref{3.1}) follows from the equality
$N_\ell=-\frac{1}{\pi}\Imag\int_0^{+\infty}\frac{{f'}_\ell(k)}{f_\ell(k)}\,dk=
-\frac{1}{\pi}\int_0^{+\infty}(\frac{d}{dk}\delta_\ell(k))\,dk$, which coincides
with formula (\ref{3.2}). The functions $f_\ell(k)$ are the so--called Jost functions
\cite{DeAlfaro}, and are related to the phase--shifts $\delta_\ell$ as follows:
$e^{2\rmi\delta_\ell(k)}=\frac{f_\ell(k)}{f_\ell(-k)}$. Therefore formulae (\ref{3.1}) and
(\ref{3.2}) hold true for the same class of potentials.}:
\beq
\frac{1}{2\pi}\int_0^{+\infty} \tau_\ell(E)\,dE + N_\ell = 0.
\label{3.2}
\eeq
Taking this formula for granted, we have a one--to--one correspondence between $\tau_\ell(E)>0$
(time delay) and $\tau_\ell(E)<0$ (time advance). In fact, suppose that $N_\ell$ is a non--negative integer number,
which depends only upon the discrete spectrum, then, in the continuum, to a
positive time delay due to a resonance it should correspond a negative time delay, i.e., a time advance,
due to an antiresonance, in a way such that the sum rule (\ref{3.2}) be satisfied (see also Ref. \cite{Wigner2}).
But equalities (\ref{3.1}) and (\ref{3.2}) are not free from ambiguities: in general the value of
$\delta_\ell(\infty)$ is unknown and, moreover, although $\delta_\ell(0)=0$ mod $\pi$, no hints of
which multiple of $\pi$ we are referring to is available.
In spite of these ambiguities the sum rule (\ref{3.2}) suggests that, in the pure elastic scattering and
in a given partial wave, resonances and antiresonances can be related in a one--to--one fashion.
In the following we shall present a theory in which resonances and antiresonances are treated
symmetrically.

With this program in mind, we start from the continuity equation, which reads:
\beq
\frac{\partial w}{\partial t} + \nabla \cdot \mathbf{j} = 2 (\Imag V_{\rm eff}) w,
\label{3.3}
\eeq
where $w=\chi^*\chi$ ($\chi$ being the wavefunction of the system), $\mathbf{j}$ is the current density:
$\mathbf{j} = \rmi\{\chi\nabla\chi^*-\chi^*\nabla\chi\}$, $V_{\rm eff}$ is the sum of the potential and of
the centrifugal barrier.
If we consider a class of local potentials which admits a continuation of the partial waves $f_\ell$
from integer values $\ell$ to complex values $\lambda$ of the angular momentum,
then $\Imag V_{\rm eff}$ can acquire values different from zero, even
if the potential is a real--valued function. In fact, from the centrifugal barrier
we have a term of the form $\Imag\left(\frac{\lambda(\lambda+1)}{2\mu R^2}\right)$,
which is different from zero if $\Imag\lambda\neq 0$. When we consider a resonance state, then there is
a continuous drain of probability from the region of interaction, in view of the
leakage of the flux of particles due to the tunnelling across the centrifugal barrier; the
probability of finding the scattered particle inside a given sphere decreases with time.
The only possibility of keeping up with this loss of probability, as required by
the continuity equation (\ref{3.3}), is to introduce a source somewhere: the source must
be proportional to $\Imag\lambda$ in order to compensate the effect of tunnelling
across the centrifugal barrier. The condition for the source to be emitting is just
$\Imag\lambda>0$ (see Ref. \cite{DeAlfaro}). Later on we shall relate this term to
an amplitude pole singularity located in the first quadrant of the CAM plane.

\smallspace

\noindent
{\bf Remark:} In the resonance interaction region
the probability of finding the scattered particles inside a given sphere decreases with time.
This decreasing behavior appears in the relation between the imaginary part of $E$,
$\Imag E = -\frac{\Gamma}{2}$, and the resonance mean life $\tau$; $\tau$
is large if the probability leakage rate is small \cite{DeAlfaro}. Since the probability of finding
the scattered particles somewhere in space is time independent, the loss of probability
must be compensated by a source, which is precisely provided by the centrifugal barrier.
However, this loss of probability should not be confused with the connection between
the time delay and the statistical density of states in scattering. In fact, as can be shown by the
Beth--Uhlenbeck formula, the density of states increases with the interaction \cite{Huang}.

\smallspace
Now, the heart of the matter is how to describe the antiresonances.
In view of the symmetry between
resonances and antiresonances, as suggested by the Levinson's sum rule, we should expect
a symmetric mechanism leading to an interpretation of the antiresonances in terms of
sinks, instead of sources. But, if the class of potentials under consideration
is restricted to the local ones, then the pole singularities of the amplitude lie in the
first quadrant of the CAM plane (i.e., $\Imag\lambda > 0$), and again we
necessarily have emitting sources \cite{DeAlfaro}. The only chance is enlarging the class of the
admitted potentials, including the non--local ones, which depend upon the angular momentum.
Now, the following question naturally arises: can this enlarged class of potentials
really describe the physical process leading to antiresonances? With this in mind,
let us return to analyze the behavior of $\delta_\ell$ ($\ell$ being fixed);
keep in mind, as a typical example, $\delta_{(\ell=2)}$ in the
$\alpha$--$\alpha$ elastic scattering (see Fig. \ref{fig_1}).
First it crosses the value $\delta_\ell=\frac{\pi}{2}$ with positive derivative, i.e.,
$\frac{d\delta_\ell(E)}{dE}>0$, and, accordingly, we observe a resonance peak in the
cross--section; next, at higher energy, but fixed $\ell\simeq kR$
($k$ is the momentum and $R$ is the interparticle distance), $\delta_\ell$ will cross
$\frac{\pi}{2}$ downward (i.e., with negative derivative $\frac{d\delta_\ell(E)}{dE}<0$),
and an antiresonance is produced. Therefore we observe antiresonances at a higher
value of the momentum $k$; it follows that the interparticle distance $R$ decreases, and the
composite structure of the interacting clusters comes out. In the specific example of the
$\alpha$--$\alpha$ elastic collision, we can say that, at the energy close to the antiresonances,
the $\alpha$--particles can no longer be described simply as bosons, but the fermionic character
of the nucleons emerges. From the Pauli's principle, and the corresponding antisymmetrization,
it derives a repulsive force, which explains the phase--shift decrease.

When the two clusters penetrate each other, Pauli exchange forces enter the game. The
nucleon--nucleon interaction which accounts for the exchange, and which is used
in the antisymmetrization process, is generally represented by a potential of
Gaussian form: $V_{p,q} \propto V_0 \exp(\cK |\mathbf{r}_p-\mathbf{r}_q|^2)[{\cC}(1+P_{pq}^r)]$,
$P_{pq}^r$ being the operator that exchanges the space coordinates of the $p^{\rm th}$
and the $q^{\rm th}$ nucleons, whose locations are represented by the vectors
$\mathbf{r}_p$ and $\mathbf{r}_q$, respectively; $\cK$ and ${\cC}$ are constants. Then, various
procedures in use, like the resonating group, the complex--generator coordinates and
the cluster coordinate methods, lead to describe the interaction between clusters
by means of an integro--differential equation of the following form:
\beq
[-\Delta+V_d]\chi(\bR)+g\int_{\R^3}V(\bR,\bR')\chi(\bR')\,d\bR'=E\chi(\bR),
\label{3.13}
\eeq
where $\hbar=2\mu=1$ ($\mu$ is the reduced mass of the clusters), $g$ is a real--valued
coupling constant, $E$, in the case of the scattering process, denotes the scattering relative
kinetic energy of the two clusters in the center of mass system, $\Delta$ is the relative motion
kinetic energy operator, and, finally, $V_d$ is the potential due to the direct forces.
Now, we assume that $V(\bR,\bR')$ is a real and symmetric function:
$V(\bR,\bR')=V^*(\bR,\bR')=V(\bR',\bR)$, and, moreover, that both the nucleon--nucleon
potentials and the wavefunctions are rotationally invariant. Then, $V(\bR,\bR')$ depends
only on the lenghts of the vectors $\bR$ and $\bR'$, and on the angle $\gamma$ between them, or
equivalently on the dimension of the triangle $(0,\bR,\bR')$, but not on its orientation. Hence,
$V(\bR,\bR')$ can be formally expanded as follows:
\beq
V(\bR,\bR')=\frac{1}{4\pi RR'}\sum_{s=0}^\infty (2s+1) V_s(R,R') P_s(\cos\gamma),
\label{3.14}
\eeq
where $\cos\gamma=\frac{\bR\cdot\bR'}{RR'}$, $R=|\bR|$, $P_s(\cdot)$ are the Legendre polynomials, and the
Fourier--Legendre coefficients are given by:
\beq
V_s(R,R')=4\pi RR'\int_{-1}^{1}V(R,R';\cos\gamma)P_s(\cos\gamma)\,d(\cos\gamma).
\label{3.15}
\eeq
Next, we expand the relative motion wavefunction $\chi(\bR)$ in the form:
\beq
\chi(\bR)=\frac{1}{R}\sum_{\ell=0}^\infty\chi_\ell(R) P_\ell(\cos\theta),
\label{3.16}
\eeq
where $\ell$ is now the relative angular momentum between the clusters. Since $\gamma$ is the
angle between the vectors $\bR$ and $\bR'$, whose directions are determined by the angles
$(\theta,\phi)$ and $(\theta',\phi')$, respectively, we have:
$\cos\gamma=\cos\theta\cos\theta'+\sin\theta\sin\theta'\cos(\phi-\phi')$.
Then, by using the following addition formula for the Legendre polynomials:
\beq
\int_0^\pi\int_0^{2\pi}P_s(\cos\gamma)P_\ell(\cos\theta')\sin\theta'\,d\theta' d\phi'=
\frac{4\pi}{2\ell+1}P_\ell(\cos\theta)\delta_{s\ell},
\label{3.17}
\eeq
from (\ref{3.13}), (\ref{3.14}), (\ref{3.16}) and (\ref{3.17}) we obtain:
\beq
{\chi''}_\ell(R)+k^2\chi_\ell(R)-\frac{\ell(\ell+1)}{R^2}\chi_\ell=
g\int_0^{+\infty}V_\ell(R,R')\chi_\ell(R')\,dR',
\label{3.18}
\eeq
where $k^2=E$ (instead of $k^2=2E$ as in formula (\ref{6})), in agreement with the position
$2\mu=1$ (see Eq. (\ref{3.13})), and the local potential is now supposed to be included into the non--local one.

In order to illustrate how a sink, instead of a source, can be obtained by introducing a non--local potential,
let us consider the following very simple model. Suppose that $V(\bR,\bR')$ can be factorized as follows:
$V(\bR,\bR')=V_*(R)\delta(R-R')v(\cos\gamma)$, ($\delta$ being the Dirac distribution).
Then, from (\ref{3.15}) the partial potentials $V_s(R,R')$ become:
\beq
V_s(R,R')=4\pi RR'V_*(R)\delta(R-R')\int_{-1}^{1}v(\cos\gamma)P_s(\cos\gamma)\,d(\cos\gamma).
\label{part-pot}
\eeq
Now, we suppose that $v(\cos\gamma)=\frac{C}{c-\cos\gamma}$ ($C,\,c$ constants, $c>1$); then, from
(\ref{part-pot}) we have (see Ref. \cite[p. 316, formula (17)]{Bateman}):
\beq
V_s(R,R')=8\pi C RR'V_*(R)\delta(R-R')Q_s(c),
\label{part-pot2}
\eeq
where $Q_s(c)$ denotes the second kind Legendre function; now, inserting Eq. (\ref{part-pot2}) into
(\ref{3.18}), and assuming suitable conditions of continuity for $V_*(R)$, the integral
in (\ref{3.18}) reduces to $V_\ell(R)\chi_\ell(R)=8\pi CR^2V_*(R)Q_\ell(c)\chi_\ell(R)$.
Next, $Q_\ell(c)$ can be regarded as the restriction of the
function $Q(\lambda;c)$ ($\lambda\in\C$) to the values $\lambda=0,1,2,\ldots$. Moreover,
$Q(\lambda;c)$ is holomorphic in the half--plane $\Real\lambda>-1$, and tends to zero uniformly
in any fixed half--plane $\Real\lambda\geqslant\delta>0$, as $|\lambda|\rightarrow\infty$.
Therefore, in view of the Carlson theorem \cite{Boas},
it represents the unique Carlsonian interpolation of the sequence $\{Q_\ell(c)\}_{\ell=0}^\infty$: i.e., the
unique analytic continuation from integer physical angular momentum $\ell$ to complex--valued
angular momentum $\lambda$. We have thus realized the unique analytic interpolation of the partial potentials
$V_\ell(R)$, given by $V(\lambda, R; c)=8\pi C_0 R^2 V_*(R) Q(\lambda; c)$ (the constant $C_0$ including
the coupling constant $g$).

Next, returning to (\ref{3.3}), and using once again the standard arguments which lead to the continuity
equation \cite{Landau}, we obtain:
\beq
\frac{\partial w}{\partial t} + \nabla \cdot \mathbf{j} = 2w (\Imag V_{\rm eff})
= 2w\left[\frac{\Imag[\lambda(\lambda+1)]}{R^2}+\Imag V(\lambda, R; c)\right].
\label{new3}
\eeq
We may thus have contributions to $\Imag V_{\rm eff}$ which derive not only from
the complexification of the centrifugal barrier but also from the potential itself. In other words,
even if the potential is a real--valued function, nevertheless it can generate (positive or negative)
contributions to $\Imag V_{\rm eff}$ in view of its angular momentum dependence.

One can then generalize\footnote{We plan to present this mathematical study elsewhere.} the model illustrated
above and find the conditions to impose on the partial potentials $V_\ell(R,R')$ in order
to obtain a unique Carlsonian interpolation $V(\lambda; R, R')$. Accordingly, one can obtain a generalization
of Eq. (\ref{new3}) which presents, in any case, a contribution to $\Imag V_{\rm eff}$ depending
on $V(\lambda; R, R')$ ($\lambda\in\C$).

In the case of local Yukawian potentials it can be proved \cite{DeAlfaro} that the locations of the
scattering amplitude singularities (poles) are necessarily restricted to the first quadrant: i.e.,
$\Imag\lambda\geqslant 0$. This is in agreement with the connection between resonances and sources
associated to the centrifugal barrier, in the sense illustrated above. Conversely, in the case
of non--local potentials the scattering amplitude singularities can lie in the first and in the fourth
quadrant; further, in connection with the antiresonances, repulsive forces become active and the probability
of finding the scattered particles within a given sphere increases with time, since an outgoing
signal can appear at a time earlier than it would have been possible in the absence of the scatterer.
Using arguments similar to those presented above for the analysis of the resonances,
we can argue that the only way of keeping up with this increase of probability, as required by the
continuity equation, is to introduce a sink: i.e., a term proportional to $\Imag\lambda<0$. As we shall
see in the next section, this corresponds to a pole singularity of the scattering amplitude in the
fourth quadrant.

\smallspace

\noindent
{\bf Remarks.}

(i) It must be neatly distinguished the difference between the analytic interpolation
of the partial potentials $V_\ell$ and that of the partial waves $a_\ell$
(which will be introduced in the next subsection).
The Carlsonian continuation of the partial potentials, which we have illustrated above, generates
a non--local potential holomorphic in $\Real\lambda>-\frac{1}{2}$ (recall, for instance, the properties
of the second kind Legendre function $Q(\lambda; c)$). Instead, the partial waves $a_\ell$ are the
restriction to integers of a function $a(\lambda; E)$ ($\lambda\in\C$, $E$ fixed) which, in the
case of local potentials of the Yukawian class, is meromorphic in the half--plane
$\Real\lambda>-\frac{1}{2}$ and holomorphic for $\Real\lambda>L-\frac{1}{2}$ ($L\geqslant 0$). In
the case of non--local potentials the geometry of the analyticity domain of $a(\lambda,E)$ changes;
in particular, it can be proved that $a(\lambda,E)$ ($E$ fixed) is meromorphic in an angular
sector\footnote{A more precise specification of the domain $\Lambda$ requires a detailed mathematical
analysis that we plan to publish in a mathematical physics journal.} $\Lambda$ contained in
the half--plane $\Real\lambda>-\frac{1}{2}$.
Resonances and antiresonances are precisely related to
the singularities of $a(\lambda,E)$ lying in the angular sector $\Lambda$ and placed in the first
and in the fourth quadrant, respectively.

(ii) In Ref. \cite{DeMicheli2} we have studied the interaction between clusters of particles
tied up by harmonic oscillators. The spectra associated with these potentials present degeneracies, which
can be removed by the action of forces depending upon the angular momentum, like
non--local potentials. Once these degeneracies have been removed, the spectra proper of the
rotational bands come out. Indeed, when the two clusters penetrate each other, the fermionic
character of the nucleons composing the clusters emerges and, accordingly, exchange forces
of the Pauli type arise. A typical example is the $\alpha$--$\alpha$
interaction discussed above, or also the $\pi^+$--p collision where, instead of the nucleons, one should
take into account the fermionic character of the quarks.
In this connection it should be pointed out that, within the limits of the present analysis,
meson--exchange forces or similar do not play here any role. \\
Finally, it is worth recalling that the minimization of functionals, in the sense of the Ritz variational calculus,
containing a Hamiltonian composed by harmonic oscillators and exchange forces gives rise
to an integro--differential equation of the form (\ref{3.13}): i.e., a Schr\"odinger equation for
non--local potentials.

(iii) In order to avoid any misleading interpretation of our analysis, we want strongly remark:
\begin{itemize}
\item[(a)] the decreasing phase--shifts, like those present in the antiresonances, are not at all a proof of
the non--elementarity of the interacting particles. Repulsion can occur, indeed, also for elementary systems.
Therefore, a one--to--one correspondence between antiresonances, time advance, and compositeness of the
interacting particles does not hold true.
\item[(b)] There exist echoes of resonances which can be explained and fitted by the hard--core
scattering, instead of by non--local potentials. This is the case of orbiting resonances in molecular
scattering (for details see Ref. \cite{DeMicheli3}).
\item[(c)] In conclusion, the present model, which uses pole singularities in the fourth quadrant of the CAM
plane for describing antiresonances and evaluating time advance, can be appropriate if and only if
Pauli exchange forces, deriving from antisymmetrization, come in play.
\end{itemize}

\section{Complex angular momentum representation of resonances and antiresonances}
\label{se:4}
\begin{figure}[t]
\begin{center}
\leavevmode
\psfig{file=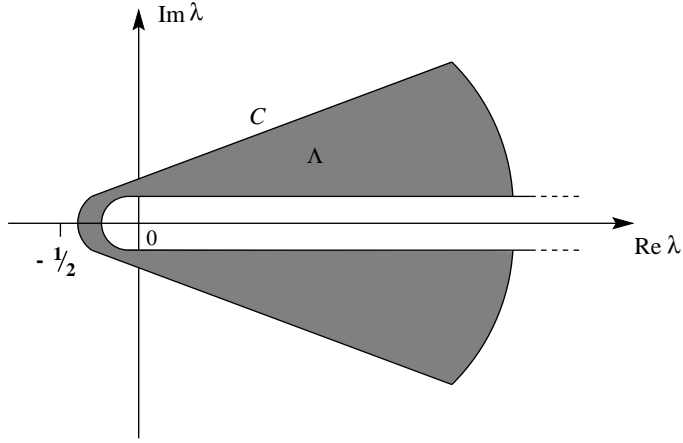,width=9cm}
\caption{\label{fig_5} \baselineskip=13pt Integration path of integral in formula (\protect\ref{4.3}).}
\end{center}
\end{figure}

Two kinds of solutions to Eq. (\ref{3.18}) must be distinguished: the scattering solutions $\chi_\ell^{(s)}(k,R)$,
and the bound--state solutions $\chi_\ell^{(b)}(R)$:
\begin{itemize}
\item[(i)] The scattering solutions satisfy the following conditions:
\begin{subequations}
\label{ele_10}
\begin{eqnarray}
&&\chi_\ell^{(s)}(k,R)=kRj_\ell(kR)+\Phi_\ell(k,R), \label{ele_10a} \\
&&\sds\Phi_\ell(k,0)=0,~~\lim_{R\rightarrow +\infty}\left[\frac{d}{dR}\Phi_\ell(k,R)-\rmi k\Phi_\ell(k,R)\right]=0,
\label{ele_10b}
\end{eqnarray}
\end{subequations}
where $j_\ell(kR)$ are the spherical Bessel functions, and the functions $d\Phi_\ell/dR$ are supposed to be
Lipschitz--continuous.
\item[(ii)]
The bound--state solutions $\chi_\ell^{(b)}(R)$ satisfy the conditions:
\beq
\int_0^{+\infty} \left | \chi_\ell^{(b)}(R) \right |^2 \, dR < \infty,~~~~~\chi_\ell^{(b)}(0)= 0.
\label{11}
\eeq
\end{itemize}
As in the case of local potentials, the asymptotic behavior of the scattering solution,
for large values of $R$, can be compared with the asymptotic behavior of the free radial function $j_\ell(kR)$,
and, correspondingly, the phase--shifts $\delta_\ell(k)$ and the scattering amplitudes
$f_\ell(k) = e^{\rmi\delta_\ell(k)}\,\sin\delta_\ell(k)$ can be defined.
Now, as in the standard collision theory, we can introduce the total scattering amplitude, which,
in view of the rotational invariance of the total Hamiltonian, can be formally expanded in terms
of Legendre polynomials as follows:
\beq
f(E,\theta)=\sum_{\ell=0}^\infty (2\ell+1)a_\ell(E)P_\ell(\cos\theta),
\label{4.1}
\eeq
where $E$ is the center of mass energy, $\theta$ is the center of mass scattering angle, $\ell$ is
the relative angular momentum of the colliding clusters, and the partial scattering amplitudes
$a_\ell(E)$ are given by:
\beq
a_\ell(E)=\frac{e^{2\rmi\delta_\ell}-1}{2\rmi k}=\frac{f_\ell}{k}~~~~~~(k^2=E,\,2\mu=\hbar=1).
\label{4.2}
\eeq
If we restrict, with suitable assumptions, the class of the admitted non--local potentials,
then we can perform a Watson--type resummation of expansion (\ref{4.1}) which yield the following
representation:
\beq
\!\!\!\!
f(E,\theta)=\frac{\rmi}{2}\int_C\frac{(2\lambda+1)a(\lambda,E)P_\lambda(-\cos\theta)}{\sin\pi\lambda}\,d\lambda
+\sum_{n=1}^N \frac{g_n(E)P_{\lambda_n}(-\cos\theta)}{\sin\pi\lambda_n},
\label{4.3}
\eeq
where the path $C$ (see Fig. \ref{fig_5}) does not necessarily has to run parallel to the imaginary axis, as
in the case of high energy physics applications. The terms $\lambda_n(E)$ give the location of the poles
of the scattering amplitude which belong to the angular sector $\Lambda$ (see Fig. \ref{fig_5}) and lying either in the
first or in the fourth quadrant of the CAM plane; $g_n(E)$ denotes the
residue of $(2\ell+1)a_\ell(E)$ at the poles. First, we consider the poles $\lambda_n=\alpha_n(E)+\rmi\beta_n(E)$
lying in the first quadrant and located within the angular sector $\Lambda$. Now, suppose
that, at a certain energy and for a specific value $n_0$ of $n$, $\alpha_{n_0}$ crosses an integer,
while $\beta_{n_0} \ll 1$: we have a pole dominance. We can thus try, in the neighborhood of a sharp and isolated
resonance, the following approximation for the amplitude:
\beq
f(E,\theta)\simeq g(E) \frac{P_\lambda(-\cos\theta)}{\sin\pi\lambda(E)},
\label{4.4}
\eeq
where we have dropped, for simplicity, the subscript $n_0$. The Legendre function $P_\lambda(-\cos\theta)$
presents a logarithmic singularity at $\theta=0$; therefore approximation (\ref{4.4})
breaks down forward, where it is needed to take into account also the contribution of the
background integral, in order to make the amplitude $f(E,\theta)$ finite and regular. Conversely,
approximation (\ref{4.4}) is satisfactory at backward angles. Further, let us note that the
divergence of the right--hand side of formula (\ref{4.4}) is of logarithmic type, then the total cross--section,
evaluated by the following integral:
\beq
\sigma_{\rm tot}=\frac{2\pi|g(E)|^2}{|\sin\pi\lambda(E)|^2}
\int_0^\pi \left|P_\lambda(-\cos\theta)\right|^2\sin\theta\,d\theta,
\label{4.5}
\eeq
converges.

Let us now project the amplitude (\ref{4.4}) on the $\ell^{\rm th}$ partial wave by means of the following formula:
\beq
\!\!\!\!
\frac{1}{2}\int_{-1}^1 P_\ell(z)\,P_\lambda(-z)\,dz = \frac{\sin\pi\lambda}{\pi(\lambda-\ell)(\lambda+\ell+1)}
~~~~(\ell=0,1,2,\ldots; \lambda\in\C).
\label{new4}
\eeq
We obtain:
\beq
a_\ell=\frac{e^{2\rmi\delta_\ell}-1}{2\rmi k}=
\frac{g}{\pi}\frac{1}{(\alpha_r+\rmi\beta_r-\ell)(\alpha_r+\rmi\beta_r+\ell+1)},
\label{4.6}
\eeq
where we write $\lambda=\alpha_r + \rmi \beta_r$ to emphasize that we are now referring to resonances. Next,
when the elastic unitarity condition can be applied, we get the following relationship among
$g$, $\alpha_r$ and $\beta_r$:
\beq
g=-\frac{\pi}{k}\beta_r(2\alpha_r+1),
\label{4.7}
\eeq
which, finally, yields:
\beq
\delta_\ell=\sin^{-1}
\frac{\beta_r(2\alpha_r+1)}{\left\{\left[(\ell-\alpha_r)^2+\beta_r^2\right]\left[(\ell+\alpha_r+1)^2+\beta_r^2\right]
\right\}^{1/2}}.
\label{4.8}
\eeq
If the colliding particles are identical, the scattering amplitude must be symmetrized (or antisymmetrized). Therefore,
instead of approximation (\ref{4.4}), we must write
\beq
f(E,\theta)\simeq \frac{g(E)}{\sin\pi\lambda(E)}\left[\frac{P_\lambda(\cos\theta)\pm P_\lambda(-\cos\theta)}
{2}\right]~~~~(0 < \theta < \pi),
\label{4.9}
\eeq
and, consequently:
\beq
\delta_\ell=\sin^{-1}\left\{\frac{1 \pm (-1)^\ell}{2}
\frac{\beta_r(2\alpha_r+1)}{\left\{\left[(\ell-\alpha_r)^2+\beta_r^2\right]\left[(\ell+\alpha_r+1)^2+\beta_r^2\right]
\right\}^{1/2}}\right\}.
\label{4.10}
\eeq
Expanding in Taylor's series the term $\lambda(E)=\alpha_r(E)+\rmi\beta_r(E)$ in a neighborhood of the
resonance energy, we can derive an estimate of the resonance width $\Gamma_r(E)$:
\beq
\Gamma_r(E) = \frac{2\beta_r\left(\frac{d\alpha_r}{dE}\right)}{\left(\frac{d\alpha_r}{dE}\right)^2+
\left(\frac{d\beta_r}{dE}\right)^2}.
\label{4.11}
\eeq
However, it must be stressed that $\Gamma_r(E)$, computed at the resonance energy $E_r$ (corresponding to
$\delta_\ell(E)=\frac{\pi}{2}$), must not be identified with the width $\Gamma$ of the observed
cross-section resonance peak.

Proceeding exactly as in the case of the resonances, we then describe the antiresonances by using poles in the
fourth quadrant of the CAM plane: $\lambda(E)=\alpha_a(E)-\rmi\beta_a(E)$ ($\beta_a>0$).
In the neighborhood of an antiresonance we have:
\beq
\delta_\ell=\sin^{-1}\left\{\frac{1 \pm (-1)^\ell}{2}
\frac{-\beta_a(2\alpha_a+1)}{\left\{\left[(\ell-\alpha_a)^2+\beta_a^2\right]\left[(\ell+\alpha_a+1)^2+\beta_a^2\right]
\right\}^{1/2}}\right\}.
\label{4.12}
\eeq
Adding the contribution of the poles lying in the first quadrant with those lying in the fourth one, we have:
\begin{eqnarray}
\delta_\ell &=& \sin^{-1}\left\{\frac{1 \pm (-1)^\ell}{2}
\frac{\beta_r(2\alpha_r+1)}{\left\{\left[(\ell-\alpha_r)^2+\beta_r^2\right]\left[(\ell+\alpha_r+1)^2+\beta_r^2\right]
\right\}^{1/2}}\right\} \nonumber \\
&+& \sin^{-1}\left\{\frac{1 \pm (-1)^\ell}{2}
\frac{-\beta_a(2\alpha_a+1)}{\left\{\left[(\ell-\alpha_a)^2+\beta_a^2\right]\left[(\ell+\alpha_a+1)^2+\beta_a^2\right]
\right\}^{1/2}}\right\}.
\label{4.13}
\end{eqnarray}
At this point the main differences between the classical Breit--Wigner theory and the present one become clear.
\begin{itemize}
\item[(i)] In the Breit--Wigner theory resonances and antiresonances are described
by means of different mathematical tools: pole singularities for the resonances, and hard--core
potential scattering for the antiresonances. The upper bound on the time advance
is rigidly fixed by the radius of the hard--core.
\item[(ii)] The resonances are not grouped in families.
\end{itemize}
Instead, the peculiar features of the present theory are:
\begin{itemize}
\item[(i')] Both resonances and antiresonances are described by pole singularities
(respectively, in the first and in the fourth quadrant of the CAM plane), which act at different
values of energy. Their effects can be separated in a rather neat way.
\item[(ii')] Resonances and antiresonances are grouped in families, and, accordingly,
the fitting parameters are constrained by the evolution of the dynamical system.
\end{itemize}

Finally, we must note that unfortunately in both theories a precise definition and evaluation of the
width $\Gamma$ of the resonances turns out to be difficult. In the Breit--Wigner theory one
introduces a parameter {\it ad hoc}, like the radius of the hard--core, whose value can change
with the energy. More generally, the width $\Gamma$ is often regarded as a function of the energy, and
various models are in use (see refs. \cite{Cutkosky,Vrana,Arndt}), as we have already explained in Section \ref{se:2}.
In the present theory, an estimate of the
{\it antiresonance width}, analogous to that given in formula (\ref{4.11}), is not available.
Then, in practice, one is naturally led to use statistical methods for evaluating the distortion
of the bell--shaped symmetry of the cross--section resonance peak.
This statistical analysis will be outlined at the end of the next section when the specific
examples of $\alpha$--$\alpha$ and $\pi^+$--p elastic scattering will be considered.

\section{Phenomenological examples}
\label{se:5}
\begin{figure}[t]
\begin{center}
\leavevmode
\psfig{file=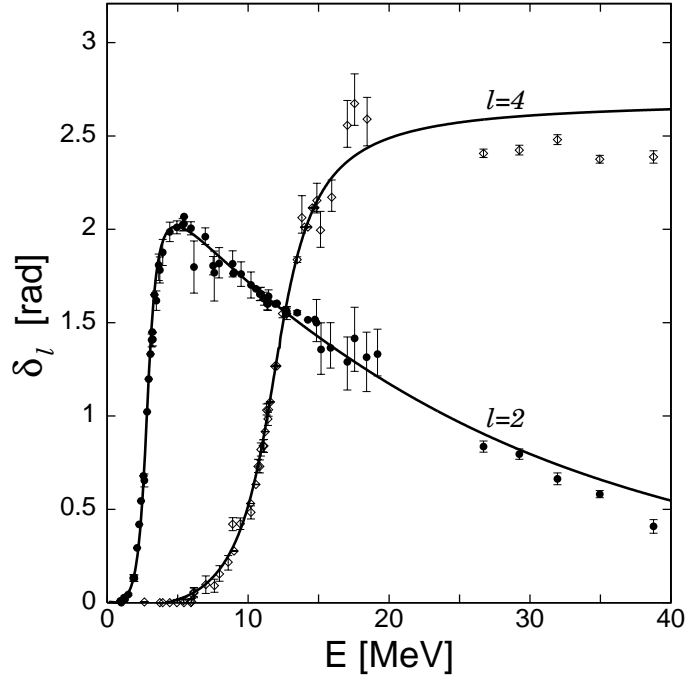,width=9cm}
\caption{\label{fig_6} \baselineskip=13pt $\alpha$--$\alpha$ elastic scattering.
Experimental phase--shifts for the partial waves $\ell=2$, $\ell=4$, and
corresponding fits accounting for both the resonance and antiresonance terms (solid lines).
For the numerical details see the legend of Fig. \protect\ref{fig_1}.}
\end{center}
\end{figure}

The theory presented above can be tested on the $\alpha$--$\alpha$ elastic scattering.
In what follows we have chosen to fit the phase--shifts, instead of the differential cross--section,
so that the action of the Coulomb interaction can be easily subtracted (see Ref. \cite{Bertero2} for
more details).
Since the $\alpha$--particles are bosons, we use the symmetrized form of formula (\ref{4.13}).
For what concerns the functions
$\alpha_r(E)$, $\beta_r(E)$, $\alpha_a(E)$ and $\beta_a(E)$, we adopt the following parametrization:
\begin{subequations}
\label{4.14}
\begin{eqnarray}
&&\alpha_r(E)[\alpha_r(E)+1] = 2IE + \alpha_0, \label{4.14a} \\
&&\beta_r(E) = b_1 E^{1/2}, \label{4.14b} \\
&&\alpha_a(E) =  a_1 E^{1/4}, \label{4.14c} \\
&&\beta_a(E) = g_0 (1-e^{-E/E_*}) + g_1 E + g_2 E^2, \label{4.14d}
\end{eqnarray}
\end{subequations}
The first equality (\ref{4.14a}) is the quantum mechanical equation of the rotator: indeed, in first
approximation, the system of two $\alpha$--particles can be viewed as a rotator whose moment of inertia
is given by $I=\mu R^2$, $\mu$ being the reduced mass and $R$ the interparticle distance. Formula
(\ref{4.14b}) states for $\beta_r(E)$ a growth which is fast for low energy, but slower for higher energy, in
agreement with the theory concerning the evolution of the resonances into surface waves for sufficiently high energy
\cite{DeMicheli1,Viano}. For what concerns $\beta_a$, the role of the exponential term is just to make, at low energy,
a smooth, though rapid, transition of $\beta_a$ from zero to $g_0$.
Unfortunately, a model which prescribes the growth properties of
$\alpha_a(E)$ and $\beta_a(E)$ is, at present, missing; it would require a more refined theory able to describe
the evolution toward semiclassical and classical phenomena. The fits of the phase--shifts are shown in Fig. \ref{fig_6}
(see also Fig. \ref{fig_1}), and the values of the fitting parameters are given in the figure legends.

Next, by using the formula:
\beq
\sigma_{\rm tot} = \frac{4\pi}{k^2}\sum_{\ell=0}^\infty(2\ell+1)\sin^2\delta_\ell,
\label{4.15}
\eeq
the total cross--section can be fitted, the fit being shown in Fig. \ref{fig_3}. The resulting fits are quite
satisfactory. Let us recall that formula (\ref{4.15}) holds true only in the energy region where the scattering is
purely elastic. In the case of $\alpha$--$\alpha$ collision this condition is satisfied up to the
energy of $17.25$ MeV \cite{Okai}. In Figs. \ref{fig_1}, \ref{fig_3} and \ref{fig_6} the bulk
of the data lie within this region. We have added some more experimental data beyond this energy to
illustrate the general trend of cross--section and phase--shifts at higher energy. Indeed, beyond $E=17.25$ MeV
formula (\ref{4.15}) can only be regarded as a drastic approximation of the total cross--section, obtained
by setting the inelasticity parameter $\eta_\ell=1$, which amounts to identify the total
cross--section with the elastic one.

Let us now consider the $\pi^+$--p elastic scattering. It presents a family of resonances
whose $J^P$ values are: $\frac{3}{2}^{+}$, $\frac{7}{2}^{+}$, $\frac{11}{2}^{+}$,
$\frac{15}{2}^{+}$, $\frac{19}{2}^{+}$. Furthermore, we also observe in the total cross--section
a resonance with $J^P=\frac{1}{2}^{-}$.
In this analysis we must take into account the spin of the proton, and
accordingly write the spin--non--flip and the spin--flip amplitudes, which read, respectively:
\begin{subequations}
\label{4.16}
\begin{eqnarray}
f(k,\theta)&=&\frac{1}{2\rmi k}\sum_{\ell=0}^\infty
\left[(\ell+1)(S_\ell^{(+)}-1)+\ell(S_\ell^{(-)}-1)\right]\,P_\ell(\cos\theta), \label{4.16a} \\
g(k,\theta)&=&\frac{1}{2k}\sum_{\ell=0}^\infty
(S_\ell^{(+)}-S_\ell^{(-)})\,P^{(1)}_\ell(\cos\theta), \label{4.16b}
\end{eqnarray}
\end{subequations}
where $P^{(1)}_\ell(\cos\theta)$ is the associated Legendre function, and
\begin{subequations}
\label{4.18}
\begin{eqnarray}
S_\ell^{(+)} = e^{2\rmi\delta_\ell^{(+)}}, \label{4.18a} \\
S_\ell^{(-)} = e^{2\rmi\delta_\ell^{(-)}}, \label{4.18b}
\end{eqnarray}
\end{subequations}
$\delta_\ell^{(\pm)}$ being the phase--shift associated with the partial wave whose total angular
momentum $J$ is equal to $\ell\pm\frac{1}{2}$.
The differential cross--section is given by
\beq
\frac{d\sigma}{d\Omega}= |f|^2 + |g|^2,
\label{4.19}
\eeq
if the proton target is unpolarized, and if the Coulomb scattering is neglected, as it is admissible at
sufficiently high energy. Next, integrating over the angles and taking into account the orthogonality of
the spherical harmonics, the following expression for the total cross--section is obtained:
\beq
\sigma_{\rm tot} = \frac{2\pi}{k^2} \sum_{J,\ell} (2J+1) \sin^2\delta_{\ell,J},
\label{4.20}
\eeq
where $J=\ell\pm\frac{1}{2}$, $\delta_{\ell,J}=\delta_{\ell,\ell\pm 1/2}$,
$\delta_{\ell,\ell+1/2}\equiv\delta_{\ell}^{(+)}$, $\delta_{\ell,\ell-1/2}\equiv\delta_{\ell}^{(-)}$.

Let us now focus on the resonance $\Delta\left(\frac{3}{2},\frac{3}{2}\right)$, which is the first
member of a family of even parity resonances.
It is generated by the interaction of a $\pi^+$ meson with a proton: the relative angular momentum
$\pi^+$--p is $\ell=1$, and the intrinsic angular momentum of the quarks in the proton is $L=0$.
Then, taking into account the spin of the proton we have: $\delta_\ell^{(+)}=\delta_{\ell,\ell+1/2}=\delta_{1,3/2}$.
The successive members of this sequence of resonances correspond to an even rotational band of the
proton states, in which the intrinsic angular momentum of the quarks in the proton is
$L=2^+,4^+,6^+,8^+$ \cite{Dalitz}. Then, the corresponding phase--shifts are
$\delta_\ell^{(+)}=\delta_{\ell,\ell+1/2}$, $\ell$ odd.
Accordingly, the antisymmetrized form of formula (\ref{4.13}) must be used:
\begin{eqnarray}
\delta_\ell^{(+)} &=& \sin^{-1}\left\{\frac{1 - (-1)^\ell}{2}
\frac{\beta_r\left(2\alpha_r+1\right)}
{\left\{\left[\left(\ell-\alpha_r\right)^2+\beta_r^2\right]\left[\left(\ell+\alpha_r+1\right)^2+\beta_r^2\right]
\right\}^{1/2}}\right\} \nonumber \\
&+& \sin^{-1}\left\{\frac{1 - (-1)^\ell}{2}
\frac{-\beta_a\left(2\alpha_a+1\right)}
{\left\{\left[\left(\ell-\alpha_a\right)^2+\beta_a^2\right]\left[\left(\ell+\alpha_a+1\right)^2+\beta_a^2\right]
\right\}^{1/2}}\right\}.
\label{4.21}
\end{eqnarray}
Analogously, the resonance $J^P=\frac{1}{2}^{-}$ is generated by the $\pi^+$--p interaction,
in which the intrinsic angular momentum of the quarks in the proton is $L=1$, and the angular momentum
of the $\pi^+$--p system is $\ell=1$ since the colliding pion is a S--wave; then for the phase--shift we have
$\delta_\ell^{(-)}=\delta_{\ell,\ell-1/2}=\delta_{1,1/2}$. This resonance is the first member of a sequence
corresponding to an odd rotational band of the proton states, in which the intrinsic angular momentum
of the quarks in the proton is $L=3^-,5^-, \ldots$. However, only the first member of this sequence gives an
observable effect to the total cross--section. Let us note that this resonance lie on a trajectory $\lambda(E)$
of the angular momentum which is different from that of the even parity family,
and then it requires an additional pole for being described.
However, in the following fits we will not account for this resonance since it lies outside
the purely elastic region of interaction.

The functions $\alpha_r$, $\beta_r$, $\alpha_a$ and $\beta_a$ are parametrized as follows:
\begin{subequations}
\label{4.23}
\begin{eqnarray}
\alpha_r &=& a_0 + a_1 (E^2 - E_0^2), \label{4.23a} \\
\beta_r &=& b_1 (E^2 - E_0^2)^{1/2} + b_2 (E^2 - E_0^2), \label{4.23b} \\
\alpha_a &=& c_0 + c_1 (E^2 - E_0^2), \label{4.23c} \\
\beta_a &=& g_0 (E^2 - E_0^2) + g_1 (E^2 - E_0^2)^2, \label{4.23d}
\end{eqnarray}
\end{subequations}
where $E$ is the energy in the center of mass frame, and $E_0$ is the rest mass of the $\pi^+$--p system.
Let us note that formula (\ref{4.23a}) differs considerably from the corresponding formula
(\ref{4.14a}). In fact, the latter refers to the non--relativistic quantum rotator, while the
former is in agreement with the phenomenological simple relationship between the total spin and the
squared mass, which, using standard notations, reads: $J\equiv\alpha(m^2)=\alpha_0+\alpha' m^2$
($\alpha_0,\,\alpha'$ constants). Let us moreover observe that from our fit (see the legend of
Fig. \ref{fig_4}) we obtain $a_1=0.9/({\rm GeV})^2$, close to the phenomenological slope
which is approximately $\alpha'\sim 1/({\rm GeV})^2$ \cite{Collins}.

Substituting the values $\delta_\ell^{(+)}$ in formula (\ref{4.20}) we can fit the total cross--section
(the data are taken from Ref. \cite{Hagiwara}), the result being shown in Fig. \ref{fig_4} (see the legend
for numerical details). It is very satisfactory, and shows with clear evidence the effect of the
antiresonance corresponding to the resonance $\Delta\left(\frac{3}{2},\frac{3}{2}\right)$.
It emerges clearly the composite structure of the interacting
particles: compare, indeed, the dashed line with the solid line in Fig. \ref{fig_4}.
The fit shown in Fig. \ref{fig_4} extends up to $p \sim 0.8 \GeV/c$. At higher impulse the elastic unitarity
condition is largely violated, and the fitting formula should be modified accordingly.
The numerical values of the parameters in Eqs. (\ref{4.23}) have been obtained by considering, in addition to
the $\Delta\left(\frac{3}{2},\frac{3}{2}\right)$, also the second member of the resonance family,
i.e., the $\frac{7}{2}^+$ resonance.
Although the latter lies outside the elastic interaction range of energy,
its mass and width can be recovered from the purely elastic cross--section (see Ref. \cite{Hagiwara}) and then
used to constrain the parameters in the fit of the even parity family of resonances.
The numerical results of this procedure have been
summarized in Table \ref{tab_1}, where also the data concerning the $\frac{11}{2}^+$ resonance are shown.

\begin{table}[t]
\caption{\label{tab_1} \baselineskip=13pt $\pi^+$--p elastic scattering.
In the present work the resonance mass is defined as the energy of the upward
$\frac{\pi}{2}$--crossing of the corresponding phase--shift $\delta_\ell(E)$.
The \textit{purely resonant} width $\Gamma_r=\Gamma_r(E)|_{E=E_r}$ (see Eq. (\ref{4.11}))
indicates the width of the resonance peak
computed without the antiresonance contribution, while the \textit{total} width $\Gamma$
stands for the width of the resonance peak accounting also for the antiresonance term.
$\frac{\Delta {\cS}}{{\cS}_{\rm Res}}$ indicates the relative increase
of skewness of the resonance peak when the antiresonance contribution
is added to the pure resonant term; here ${\cS}$ is evaluated by means of
${\cS}_{\rm stat}$ (see text).
}
{\scriptsize
\begin{tabular}{cccccccc}
\\ \hline
Name & $J^P$ & Mass [MeV] & Mass [MeV] & $\Gamma_r$ [MeV] & $\Gamma$ [MeV] &
$\Gamma$ [MeV] & $\frac{\Delta{\cS}}{{\cS}_{\rm Res}}$ \\
&& (present work) & (Ref. \protect\cite{Hagiwara}) & Purely resonant & Total &
(Ref. \protect\cite{Hagiwara}) & \\ \hline
$\Delta(1232)$ & $\frac{3}{2}^+$ & 1232.8 & $1230-1234$ & $93$ & 115 & $115-125$ & 4.1 \\
$\Delta(1950)$ & $\frac{7}{2}^+$ & 1951 & $1940-1960$ & $293$ & --- & $290-350$ & --- \\
$\Delta(2420)$ & $\frac{11}{2}^+$ & 2463 & $2300 - 2500$ & $397$ & --- & $300 - 500$ & --- \\
\hline
\end{tabular}
}
\end{table}

\begin{figure}[t]
\begin{center}
\leavevmode
\psfig{file=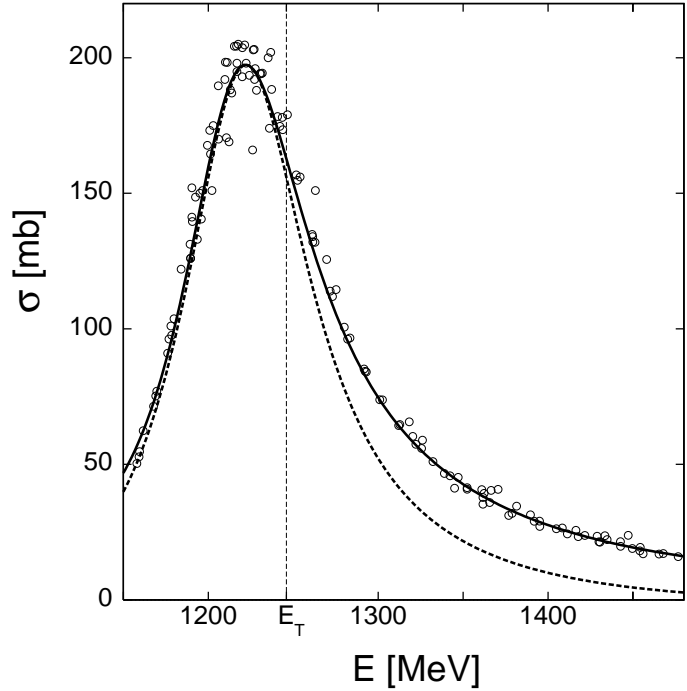,width=9cm}
\caption{\label{fig_7} \baselineskip=13pt $\pi^+$--p elastic scattering: $\Delta(\frac{3}{2},\frac{3}{2})$ resonance.
Comparison between the total cross--section computed with
(solid line) and without (dashed line) the antiresonance term. $E_T$ indicates the
energy at which the antiresonance effect is expected to set in (see text). For better comparison,
the dashed curve has been slightly translated in order to have the peaks of the two curves coincident.}
\end{center}
\end{figure}

With regard to the $\Delta\left(\frac{3}{2},\frac{3}{2}\right)$ resonance, a simple semiclassical
argument can support the interpretation of the distortion effect of the bell--shaped resonance peak in terms of
the composite structure
of the colliding particles. If we denote by $R$ the distance between the pion and the proton, supposed at rest, then
the impulse of the projectile, i.e., the pion, in the laboratory frame is $p_{\rm LAB}^\pi \sim \sqrt{2}\hbar / R$,
since $\ell = 1$. Now, setting $R$ as the distance at which the two
particles {\it get in contact} (or, in other words, setting $R$ as the interaction radius),
which is of the order of the proton radius \cite{Manley}, then the corresponding $p_{\rm LAB}^\pi$
gives an estimate of the least pion impulse at which the internal structure of the two particles enter the game in the
collision process. By using $R \sim R_{\rm proton} = 0.87$ fm \cite{Hagiwara}, we obtain
$p_{\rm LAB}^\pi \sim 320 \MeV/c$, which corresponds to the energy in the center of mass frame:
$E_T \sim 1246 \MeV$. Fig.\ref{fig_7} compares the total $\pi^+$--p cross section computed with
(solid line) and without (dashed line) the antiresonance contribution: it is clear that the antiresonance
effect sets in at an energy very close to $E_T$.

The above analysis suggests a procedure for evaluating the width $\Gamma$ of the resonance
$\Delta\left(\frac{3}{2},\frac{3}{2}\right)$.
We note, first of all, that the effect of the antiresonance is a small perturbation
to the pure resonance; this means that the reference baseline of the pure resonance distribution and that of the
observed cross--section (comprising both resonance and antiresonance) coincide within a
good approximation. Then, from the plot of the cross--section generated by only the pure resonant
term we can recover the reference baseline of this almost symmetric bell--shaped distribution by
equating its second central moment to the value of $\Gamma_r$ evaluated by means of
formula (\ref{4.11}). Next, keeping fixed this baseline, we evaluate the second central moment
of the distribution which fits the experimental cross--section peak (i.e., accounting also
for the antiresonant term). We can take as an estimate of the total width $\Gamma$ the value of this
second moment. As a measure of the degree of asymmetry of the resonance peak, which can be ascribed
to the composite structure of the interacting particles, we take the statistical skewness
$\cS_{\rm stat}$ of the distribution,
defined as $\cS_{\rm stat}=\mu_3/\mu_2^{3/2}$, where $\mu_2$ and $\mu_3$ are the second and third central
moments of the distribution (see Table \ref{tab_1}).

If the asymmetry of the bell--shaped resonance peak is very large, as in the case of
the $\alpha$--$\alpha$ elastic scattering (see Fig. \ref{fig_3}), and the antiresonance effect cannot
be regarded as a small perturbation to the pure resonance, then the baseline of the pure
resonance peak and that of the observed experimental cross--section differ significantly.
The statistical method illustrated above can no longer be applied, and we are forced to
follow a more pragmatic attitude. Since the symmetry due to the antiresonance effect
sets in just after the resonance maximum, we can regard $\Gamma$ as the half--width
at half--maximum of the resonance peak, like in the Breit--Wigner theory. From the plot
of the pure resonance cross--section we obtain a bell--shaped distribution whose
full--width at half--maximum agrees with the value of $\Gamma_r$ evaluated by
formula (\ref{4.11}). Then we can give an estimate of the total width $\Gamma$
by evaluating the full--width at half--maximum of the curve fitting the experimental
cross--section (including both resonance and antiresonance terms). In this case the
asymmetry of the resonance peak can be estimated by means of a {\it phenomenological
skewness} $\cS_{\rm phen}$, defined as follows:
first we compute the difference between the two half--maximum semi--widths, measured with
respect to the energy of resonance $E_r$; then we define the degree of asymmetry $\cS_{\rm phen}$
as the ratio between this value and the full--width $\Gamma$.

\end{document}